\documentclass[aps,pra,superscriptaddress,twocolumn,a4paper,reprint,eprint,longbibliography,footinbib]{revtex4-1}
\usepackage[utf8]{inputenc}
\usepackage{graphicx}
\usepackage{amsmath}
\usepackage{amsfonts}
\usepackage{amssymb}
\usepackage{bbm}
\usepackage{bm}
\usepackage{bbold}
\usepackage[usenames,dvipsnames,svgnames,table]{xcolor}
\usepackage[USenglish]{babel}
\usepackage{hyperref}
\usepackage{grffile}

\newcommand{\eq}[1]{Eq.~(\ref{#1})} 
\newcommand{\fig}[1]{Fig.~\ref{#1}} 
\newcommand{\ignore}[1]{}

\allowdisplaybreaks

 \pdfpageattr{/Group <</S /Transparency /I true /CS /DeviceRGB>>} 

\begin{document}

\title{Robustness of unconventional $s$-wave superconducting states
  against disorder}

\author{D. C. Cavanagh}
\email{david.cavanagh@otago.ac.nz}
\affiliation{Department of Physics, University of Otago, P.O. Box
56, Dunedin 9054, New Zealand}
\author{P. M. R. Brydon}
\email{philip.brydon@otago.ac.nz}
\affiliation{Department of Physics and MacDiarmid Institute for
Advanced Materials and Nanotechnology, University of Otago, P.O. Box
56, Dunedin 9054, New Zealand}

\date{January 14, 2020}

\begin{abstract}
We investigate the robustness against disorder of superconductivity in
multiband systems where the fermions have four internal degrees of
freedom. This permits unconventional $s$-wave
pairing states, 
which may transform nontrivially under crystal symmetries. 
Using the self-consistent Born approximation, we develop a general
theory for the effect of impurities on the critical
  temperature, and find that the 
presence of these novel $s$-wave channels significantly modifies
the conclusions of single-band theories. We
apply our theory to two candidate topological superconductors, YPtBi
and Cu$_x$Bi$_2$Se$_3$, and show that the novel $s$-wave states
display an enhanced resilience against disorder, which extends to
momentum-dependent pairing states with the same crystal symmetry. The
robustness of the $s$-wave states can be quantified in terms of their
superconducting fitness, which can be readily evaluated for model systems.
\end{abstract}

\maketitle

\section{Introduction}

It is a textbook result that
the critical temperature $T_c$ of 
a conventional $s$-wave spin-singlet superconductor is insensitive to
nonmagnetic disorder~\cite{Mineev1999}. This is a consequence of Anderson's
theorem~\cite{Anderson1959}: since this state
has an isotropic gap and pairs electrons in time-reversed partner
states, there is no depairing effect from the time-reversal-invariant
scattering off the impurities. On the other hand, the sign-reversing gaps
of unconventional superconductors are averaged to
zero by the impurity-scattering across the Fermi surface, and these
pairing states are suppressed by weak disorder with normal state
 elastic scattering rate (SR) $\tau^{-1} \sim k_BT_c$. 

Recently there has been much interest in $s$-wave pairing states which
\emph{do not} pair time-reversed partner states~\cite{Ong2016,Vafek2016,AgterbergFeSe2017,Kawakami2018,Oiwa2018,Moeckli2018,FuBerg2010,Brydon2016}. This can
occur in systems where the electrons  have additional discrete
degrees of freedom, such as orbital or
sublattice indices. These permit novel ways to satisfy the fermionic
antisymmetry of  
the Cooper pair wavefunction in a relative $s$-wave, e.g. a spin-triplet
orbital-singlet state. Such pairing states typically belong to a
nontrivial irreducible representation (irrep) of the point group. They
have been proposed in a variety of
materials~\cite{Ong2016,Vafek2016,AgterbergFeSe2017,Kawakami2018,Oiwa2018,Moeckli2018}, 
but here we focus on 
Cu$_x$Bi$_2$Se$_3$~\cite{FuBerg2010} and 
YPtBi~\cite{Brydon2016}. Experiments indicate a  fully-gapped
nematic superconducting state in
Cu$_x$Bi$_2$Se$_3$~\cite{Matano2016,Yonezawa2016,Tao2018}, which
naturally arises from a time-reversal-invariant combination of the
odd-parity $s$-wave states in 
the $E_u$ irrep~\cite{Fu2014}. In YPtBi there is evidence 
of a nodal superconducting gap~\cite{Kim2018}, which could be
explained by a time-reversal symmetry-breaking combination of
even-parity $s$-wave states which support exotic Bogoliubov Fermi
surfaces~\cite{Brydon2016,Timm2017}. 

Since the novel $s$-wave states do not pair time-reversed
  partners, Anderson's theorem does 
not apply and we may expect them to be
highly sensitive to disorder. Indeed, expressed in a psuedospin
 band basis, the novel  $s$-wave states show a sign-changing 
gap, which averages to zero across the Fermi
surface~\cite{Yip2013,Brydon2016}. However, since the impurity 
potential in the pseudospin  band basis may be anisotropic, the
\emph{impurity-averaged} gap may not 
vanish, which can lead to unconventional impurity
effects~\cite{Fomin2018,Eltsov2019}. This anisotropy naturally
  appears when the states at the Fermi surface have a strong spin-orbital
  texture. Indeed, it was   
it was shown in 
Ref.~\cite{MichaeliFu2012} that the spin-orbital texture of the
  electronic states at the Fermi surface in Cu$_x$Bi$_2$Se$_3$
  generates such an anisotropy, granting the
novel $s$-wave
$A_{1u}$ state enhanced robustness against
disorder. It is nevertheless unclear if a general principle underlies
this result, or if it applies to other pairing states. 

In this paper we use the self-consistent Born approximation to study 
the effect of disorder on the critical temperature of a
superconducting state in a system where the fermions have four degrees
of freedom. In Sec.~\ref{sec:generaltheory} we develop a general framework which explicitly keeps
track of
these degrees of freedom, naturally generalizing the familiar
results of single-band theories with disorder~\cite{Mineev1999}.
Our computationally-straightforward approach generalizes
    and extends earlier 
    works~\cite{MichaeliFu2012,Nagai2015,Scheurer2016}, and can be 
    readily  applied to new materials.
As concrete examples, in Secs.~\ref{sec:YPtBi} and~\ref{sec:CuxBi2Se3} we apply our theory to YPtBi and
Cu$_x$Bi$_2$Se$_3$, respectively. 
We find that  
nontrivial $s$-wave states proposed for these systems show a parametrically-enhanced
robustness against  
disorder, which is shared with other states in the same irrep
according to their similarity to the $s$-wave states at the
Fermi surface.  
In the discussion of Sec.~\ref{sec:discussion}, we show that the robustness of the $s$-wave states is quantified in terms of the
superconducting fitness~\cite{Ramires2016,Ramires2018}, which can be
readily evaluated for 
model Hamiltonians. Although robust unconventional states are generally
possible, systems with nontrivial inversion operator
are particularly favourable.

\section{General theory}\label{sec:generaltheory}

Our starting point is a generic model of a
fermionic system with four 
internal degrees of freedom that is invariant under
time reversal and inversion. The normal-state
Hamiltonian is $H = \sum_{\bm k}c^\dagger_{\bm k}{\cal H}_{\bf
  k}c_{\bm k}$, 
where $c_{\bm k}$ is a four component spinor encoding the 
internal degrees of freedom, and  the matrix 
${\cal H}_{\bm
  k}$ has the general form~\cite{Brydon2018b},
\begin{equation}
 {\cal H}_{\bm
  k} = \epsilon_{{\bm k},0}\,\mathbb{1}_4 +
\vec{\epsilon}_{\bm k}\cdot\vec{\gamma}\,, \label{eq:genH}
\end{equation}
where $\mathbb{1}_4$ is the $4\times4$ unit matrix and 
$\vec{\gamma}=(\gamma^1,\gamma^2,\gamma^3,\gamma^4,\gamma^5)$ is the
vector of the five mutually-anticommuting Euclidean Dirac matrices.
The real functions
$\epsilon_{{\bm k},0}$ and $\vec{\epsilon}_{\bm k}
  =(\epsilon_{{\bm k},1},\epsilon_{{\bm k},2},\epsilon_{{\bm k},3},
  \epsilon_{{\bm k},4},\epsilon_{{\bm k},5})$
are the coefficients of these  
matrices. The Hamiltonian in \eq{eq:genH} has the doubly degenerate
eigenvalues $E_{{\bm k},\pm}=\epsilon_{{\bm k},0} \pm
|\vec{\epsilon}_{\bm k}|$. The internal degrees of freedom can either
transform trivially (${\cal I}=\mathbb{1}_4$) or nontrivially (${\cal
  I}=\gamma^1$) under inversion. The time-reversal operator is
  ${\cal T}=U_T{\cal K}$, where ${\cal K}$ is complex conjugation and 
the unitary part can be  expressed in terms of the Euclidean
  Dirac matrices without loss of generality as $U_T=\gamma^3\gamma^5$.

    The pairing potential for a general superconducting state is $\Delta_{\bm k} =
    \Delta_0\tilde{\Delta}_{\bm k}$ where  $\Delta_0$ is the
      magnitude and
    \begin{equation}
      \tilde{\Delta}_{\bm k} = f_{{\bm k}}\gamma^{\alpha}\gamma^{\beta}U_T \,. \label{eq:genDelta}
    \end{equation}
 Here $f_{\bm{k}}$ is a normalised form factor,
    chosen such that fermionic antisymmetry
    $\tilde{\Delta}_{\bm k}=-\tilde{\Delta}^T_{-{\bm k}}$ is satisfied. Because the
    pairing potential $\tilde{\Delta}_{\bm k}$ is a $4\times4$ matrix, there
    are six terms in~\eq{eq:genDelta} for which an $s$-wave form
    factor (i.e. $f_{\bm k}=1$)
    is permitted by fermionic antisymmetry. This is always possible for $\alpha=\beta=0$ (where $\gamma^0=\mathbb{1}_4$), which
    describes pairing between electrons in 
    time-reversed partner states, and hence generalizes the
    usual $s$-wave spin-singlet state. The five other 
    channels where an $s$-wave form-factor is allowed
      have a nontrivial dependence on the internal degrees of freedom,
      where $\alpha$ and $\beta$ in~\eq{eq:genDelta} are different
    and not both zero. These additional $s$-wave channels
  typically belong to nontrivial irreps.
    
The nontrivial $s$-wave channels do not
    generally pair electrons in time-reversed partner states,
    and hence typically involve both intraband and interband
    pairing. 
To quantify
    	the degree of interband pairing for a pairing state $\tilde{\Delta}_{\bm
    		k}$ at wavevector ${\bm k}$, Ref.~\cite{Ramires2018} introduced the quantity  
    	$F_C({\bm k}) =
    	\frac{1}{4}\text{Tr}\{|{\cal H}_{\bm k}\tilde{\Delta}_{\bm
    		k}-\tilde{\Delta}_{\bm
    		k}{\cal H}_{-\bm k}^T|^2\}$, where ${\cal H}_{\bm k}\tilde{\Delta}_{\bm
    		k}-\tilde{\Delta}_{\bm
    		k}{\cal H}_{-\bm k}^T$ is referred to as the ``superconducting
    	fitness''~\cite{Ramires2016} and is 
    	vanishing if there is no interband pairing.
The superconducting
                    fitness also controls the form of the
                      superconducting gap in the low-energy
                      spectrum. Specifically, the $s$-wave states
                      (i.e. $\tilde{\Delta}_{\bm k}=\tilde{\Delta}$) open
                      a gap of magnitude \cite{Ramires2019} 
    \begin{equation}
    \Delta_0\sqrt{1-\tilde{F}_{C}(\bm k)},
    \end{equation}
    where $\tilde{F}_{C}(\bm k) =
     4F_{C}(\bm k)/|\vec{\epsilon}_{\bm 
    	k}|^2\text{Tr}\{\tilde{\Delta}\tilde{\Delta}^\dagger\}$ is normalized such that  $\tilde{F}_C(\bm k)\leq
      1$. If $\tilde{F}_C(\bm k)=1$, there is
        no intraband pairing, and so the $s$-wave states must
        necessarily display a gap
        node. Since the spin-singlet
        analogue state is perfectly fit (i.e. $\tilde{F}_C(\bm k)=0$), it hence opens a full
        gap and there is no interband pairing, as anticipated by the fact that it pairs time-reversed
        partners. In contrast, the
        nontrivial matrix structure of the anomalous $s$-wave states
        typically results in a nonzero fitness and possibly the
        formation of  nodes.
    
    We consider isotropic scattering off potential impurities distributed
    randomly at positions ${\bm r}_j$,  described by the Hamiltonian
    \begin{equation}
      H_{\text{imp}} = \frac{V}{\Omega}\sum_{j}\sum_{{\bm k},{\bm
          k}'}e^{i({\bm k}'-{\bm k})\cdot{\bm r}_j}c^{\dagger}_{{\bm
          k}}c_{{\bm k}'} \label{eq:impHam}
    \end{equation}
    where $V$ is the impurity potential and $\Omega$ is the
    volume.
    We restrict ourselves here to the use of a scattering
      potential that is isotropic in the spin and orbital indices, as
      is the standard approach for nonmagnetic impurities
      \cite{MichaeliFu2012,Mineev1999,Nagai2015}. Although more
      complicated impurity
      potentials are possible in systems with orbital degrees of
        freedom~\cite{Scheurer2015,Moeckli2018}, our intention here is to
      understand the relationship between the spin-orbital
        texture of the normal-state bands and the
        robustness of the superconducting state to
        disorder. To this end, we focus on the simplest possible
      scattering potential in the spin-orbital basis. This
      simplification does not imply that intra- and interband
      scattering processes are equivalent, however, as such
      processes 
        depend on  matrix elements introduced by the
      transformation to the band basis. 
    Within the self-consistent
    Born approximation, the Green's 
    functions of the disordered system are
    \begin{equation}
      \bar{G}({\bm k},i\omega_n) = \sum_{j=\pm}\frac{1}{i\tilde{\omega}_{n,j} - E_{{\bm
    k},j}}{\cal
  P}_{{\bm k},j} \label{eq:GF}
    \end{equation}
    where ${\cal
  P}_{{\bm k},\pm} = \frac{1}{2}\left(\mathbb{1}_4 
    \pm\hat{\epsilon}_{\bm k}\cdot\vec{\gamma}\right)$ projects into the
    $\pm$ band at momentum ${\bm k}$ and
    $\hat{\epsilon}_{\bm k} = 
    \vec{\epsilon}_{\bm k}/|\vec{\epsilon}_{\bm k}|$. The effect of
    impurities is accounted for in the renormalized Matsubara frequencies
    $\tilde{\omega}_{n,j} = \omega_n - (2\tau_{{\bm k},j})^{-1}\text{sgn}(\omega_n)$,
    where the SR in band $j$ is 
\begin{equation}
\frac{1}{\tau_{{\bm k},j}} = \pi n_{\text{imp}} V^2 \sum_{m=\pm}{\cal N}_{m}\left(1 +
jm\hat{\epsilon}_{\bm k}\cdot\langle \hat{\epsilon}_{\bm k}\rangle_{\text{FS},m}\right)\,.\label{eq:NormSR}
\end{equation}
Here $n_{\text{imp}}$ is the concentration of impurities, ${\cal
  N}_{m}$ is the density of states of band $m=\pm$ at the Fermi 
surface, and $\langle\ldots\rangle_{\text{FS},m}$ denotes the average over
the Fermi surface of this band. The second term in the
  parentheses of~\eq{eq:NormSR} is an additional contribution to the
  scattering rate  which arises from a net average
  polarization in the internal degrees of freedom on the (single band) Fermi surface. In the following we will assume a weak
momentum-dependence of the SR and replace $\tau^{-1}_{{\bf k},j}$ by
its Fermi surface average in~\eq{eq:GF}. 

\begin{figure}
	\includegraphics[width=\columnwidth]{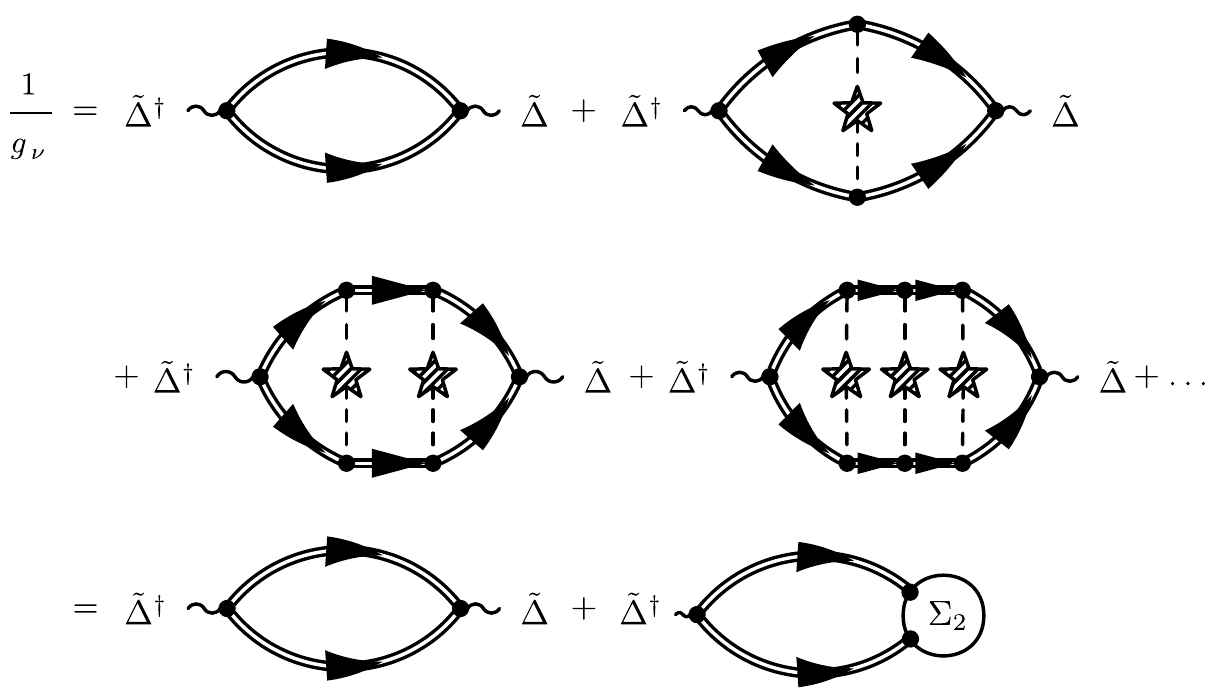}
	\caption{Diagammatic form of the linearized gap equation, taking Cooperon ladder diagrams into account. The dotted line represents the interaction with the impurity, denoted by the star, and the double line is the Green's function dressed by interactions with the impurity via the normal self-energy.}\label{fig:SCGE}
\end{figure}

\begin{widetext}
The critical temperature in the presence of disorder can be determined from the lowest-order terms of the Ginzburg-Landau free energy in the Born approximation, expanded in powers of the gap \cite{Scheurer2015,Mineev2007},
\begin{equation}
\mathcal{F}_2 = \frac{\left|\Delta_0\right|^2}{g_{\nu}}  + \frac{1}{2\beta}\sum_{i\omega_n}\int\!\!\frac{d^3k}{(2\pi)^3}\mbox{Tr}\{\Delta_{\bm k}^{\dagger}\bar{G}({\bm
  k},i\omega_n)(\Delta_{\bm k}+\Delta_0\Sigma_2)\bar{G}_{h}({\bm
    k},i\omega_n)\}
\end{equation}
\end{widetext}
where $\bar{G}_h({\bm k},i\omega_n) = \bar{G}^{T}(-{\bm k},i\omega_n)$
is the Green's function for the holes, $g_{\nu}<0$ is the attractive
interaction in a particular superconducting channel $\nu$, and $\Sigma_2$ is the anomalous
self-energy due to the impurity scattering. The critical temperature is found by minimizing the free energy with respect to $\Delta^\ast$. Cancelling an overall factor of the gap magnitude $\Delta_0$ gives an expression for the linearized gap equation 
\begin{equation}
  \frac{1}{g_{\nu}} = \frac{1}{2\beta}\sum_{i\omega_n}\int\!\!\frac{d^3k}{(2\pi)^3}\mbox{Tr}\{\tilde{\Delta}_{\bm k}^{\dagger}\bar{G}({\bm
  k},i\omega_n)(\tilde{\Delta}_{\bm k}+\Sigma_2)\bar{G}_{h}({\bm
    k},i\omega_n)\} \label{eq:linearized}
\end{equation}
which includes the Cooperon ladder diagrams (see~\fig{fig:SCGE}) via the anomalous self-energy.
 

\begin{figure}
	\includegraphics[width=\columnwidth]{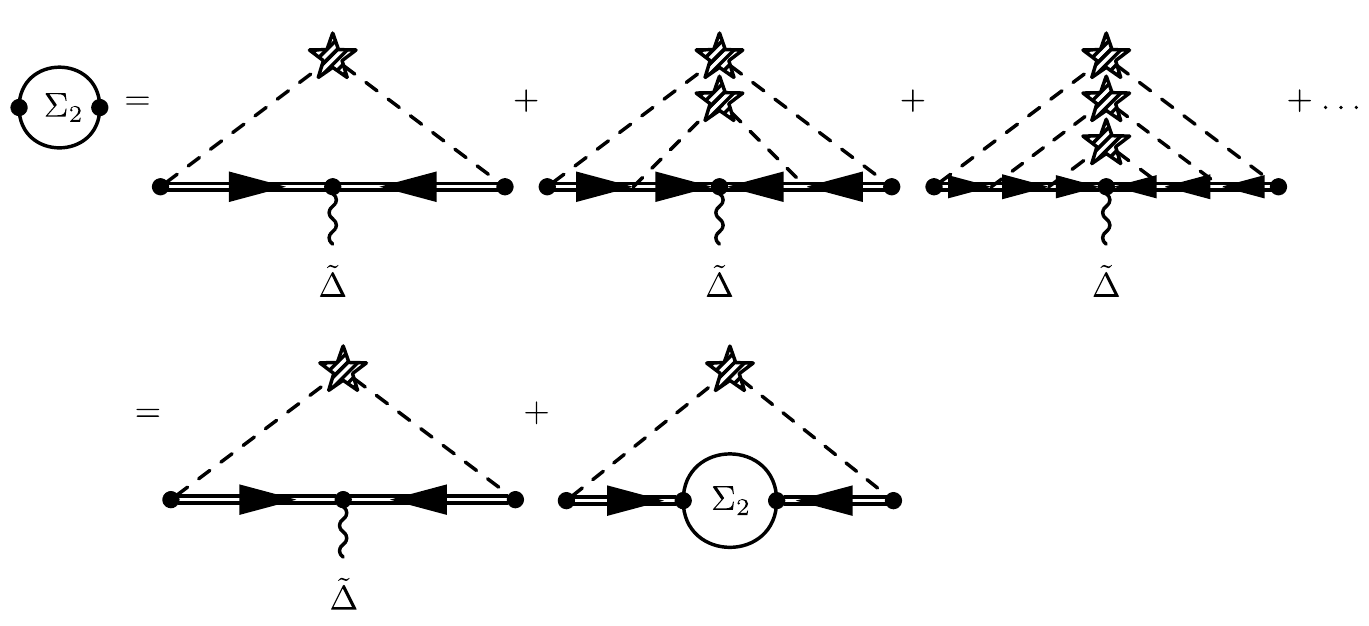}
	\caption{ Diagammatic form of the anomalous self-energy in the self-consistent Born approximation. }\label{fig:SCBA}
\end{figure}

The self-energy obeys the
self-consistency equation defined diagrammatically in~\fig{fig:SCBA},
\begin{equation}
\Sigma_2 = 
-n_{\text{imp}}V^2\int\!\!\frac{d^3k}{(2\pi)^3}\bar{G}({\bm
	k},i\omega_n)(\tilde{\Delta}_{\bm k}+\Sigma_2)\bar{G}_{h}({\bm k},i\omega_n)\,.\label{eq:Sigma2}
\end{equation}
The anomalous self-energy vanishes unless the lowest-order
contribution is nonzero: 
\begin{align}
  \Sigma^{(0)}_{2} &=  -n_{\text{imp}}V^2\int\!\!\frac{d^3k}{(2\pi)^3}\bar{G}({\bm
  k},i\omega_n)\tilde{\Delta}_{\bm k}\bar{G}_{h}({\bm k},i\omega_n)
  \notag \\
  &=  \pi n_{\text{imp}}V^2\sum\limits_{j=\pm}
  \frac{{\cal N}_j}{\left|\tilde{\omega}_{n,j}\right|} \langle {\cal P}_{{\bm
      k},j}\tilde{\Delta}_{\bm k}{\cal P}^T_{-{\bm
      k},j}\rangle_{\text{FS},j},
\ignore{       {\Big
    \{}\langle\Delta_{\bm k}\rangle_{\text{FS},j}
  + \sum\limits_{l} \lambda_l \langle\hat{\epsilon}^2_{{\bm
      k},l}\Delta_{\bm k}\rangle_{\text{FS},j} \notag \\
  & + j \sum\limits_{l} 
  \langle \hat{\epsilon}_{{\bm k},l}\left\lbrace \gamma_l,\Delta_{\bm
    k}\right\rbrace\rangle_{\text{FS},j}  \notag \\
  & +\sum\limits_{l \neq m}
  \langle \hat{\epsilon}_{{\bm k},l}\gamma_l\Delta_{\bm
    k}\gamma_m\hat{\epsilon}_{{\bm k},m}\rangle_{\text{FS},j}{\Big
    \}}} \label{eq:Sigma20}
\end{align}
where, in the final line, we have made the assumption that the
  bands are well separated and interband contributions to the
  self-energy are therefore small and can be neglected. Equation~\eq{eq:Sigma20} is the central result of our
analysis. Because of the nontrivial form of the projection operators,
the Fermi-surface average will not necessarily vanish for an
unconventional state. Although our theory has been developed for a
two-band model, this result readily generalizes to an
arbitrary number of bands.  For the two-band system considered here,
explicitly evaluating~\eq{eq:Sigma20} for  the general pairing
state~\eq{eq:genDelta} yields 
\begin{widetext}
\begin{equation}
\Sigma^{(0)}_{2}  =  \pi n_{\text{imp}}V^2\sum_{j=\pm}
\frac{{\cal N}_j}{4\left|\tilde{\omega}_{n,j}\right|} \left[\langle
                    f_{\bm k}\rangle_{\text{FS},j}\gamma^\alpha\gamma^\beta
                    + j\sum_{l=1}^{5} \langle f_{\bm
                    k}\hat{\epsilon}_{\bm{k},l}\rangle_{\text{FS},j} \left\lbrace
                    \gamma^\alpha\gamma^\beta,\gamma^l\right\rbrace  +
                    \sum_{l,m=1}^{5}\langle f_{\bm
                    k}\hat{\epsilon}_{\bm{k},l}\hat{\epsilon}_{\bm{k},m}\rangle_{\text{FS},j}\gamma^l\gamma^\alpha\gamma^\beta\gamma^m\right]U_T\,. \label{eq:s-wSig20}
\end{equation}
\end{widetext}
Due to the Fermi surface averages of the form-factor $f_{\bm k}$ with
the coefficients of the $\gamma$ matrices in the
Hamiltonian~\eq{eq:genH}, the self-energy may be nonzero even for
nontrivial form factors. 
Moreover, we observe that since $\Sigma_2^{(0)}$ (and hence
$\Sigma_2$) is
independent of momentum, it must belong
to one of the $s$-wave channels, 
and can thus be nonzero for any state in the same irrep. 
This modifies the solution of~\eq{eq:linearized} such 
that these states acquire some protection against the
disorder. This 
represents the crucial difference to the single-band case, where the
trivial form of the projection operators implies that the
anomalous self-energy vanishes for any state with a sign-changing gap, and
the critical temperature of these states is suppressed in a
universal fashion~\cite{Mineev1999}.

Although our theory applies to a general two-band system, the
analysis of systems with
multiple Fermi surfaces is complicated.  To more clearly
    reveal the universal physics due to the spin-orbital texture, therefore, in the following
    we study two examples of the simpler case
    where only one of the bands intersects the Fermi energy. 

\section{Application to $\text{YPtBi}$}\label{sec:YPtBi}

YPtBi is a zero-band-gap semimetal, where the states
close to the Fermi energy belong to the
$\Gamma_8$ band. Ignoring a weak antisymmetric spin-orbit coupling due
to the broken inversion symmetry~\cite{Brydon2016}, this is described by  
the Luttinger-Kohn model for the $j=\frac{3}{2}$ states in a cubic
material 
\begin{equation}
H = (\alpha |{\bm k}|^2 - \mu)\mathbb{1}_4 + \beta_1\sum_{i}k_i^2J_{i}^2
+ \beta_2\sum_{i\neq i^\prime}k_ik_{i^\prime}J_{i}J_{i^\prime}\,, \label{eq:LKHam}
\end{equation}
where $i$ and $i^\prime$ enumerate the Cartesian coordinates.
The $j=\frac{3}{2}$ internal angular momentum of the electrons
constitutes the four degrees of freedom in our general
model, and the $\gamma$ matrices in~\eq{eq:genH} can be parameterized as
$\vec{\gamma} = (\frac{1}{\sqrt{3}}(J_x^2-J_y^2), \allowbreak
\frac{1}{3}(2J_z^2-J_x^2-J_y^2), \allowbreak
\frac{1}{\sqrt{3}}\{J_y,J_z\}, \allowbreak
\frac{1}{\sqrt{3}}\{J_x,J_y\}, \allowbreak \frac{1}{\sqrt{3}}\{J_x,J_z\})$
with $\vec{\epsilon}_{\bm k} =
(\sqrt{3}\beta_1(k_x^2-k_y^2)/2, \allowbreak \beta_1(3k_z^2-|{\bm
  k}|^2)/2, \allowbreak \sqrt{3}\beta_2k_yk_z, \allowbreak
\sqrt{3}\beta_2k_xk_y, \allowbreak\sqrt{3}\beta_2k_xk_z)$.
The $j=\frac{3}{2}$ index transforms trivially under inversion. Experiments
show hole-like carriers in YPtBi, and so we set the chemical
potential to lie in the lower band.

\begin{figure}
  \includegraphics[width=\columnwidth]{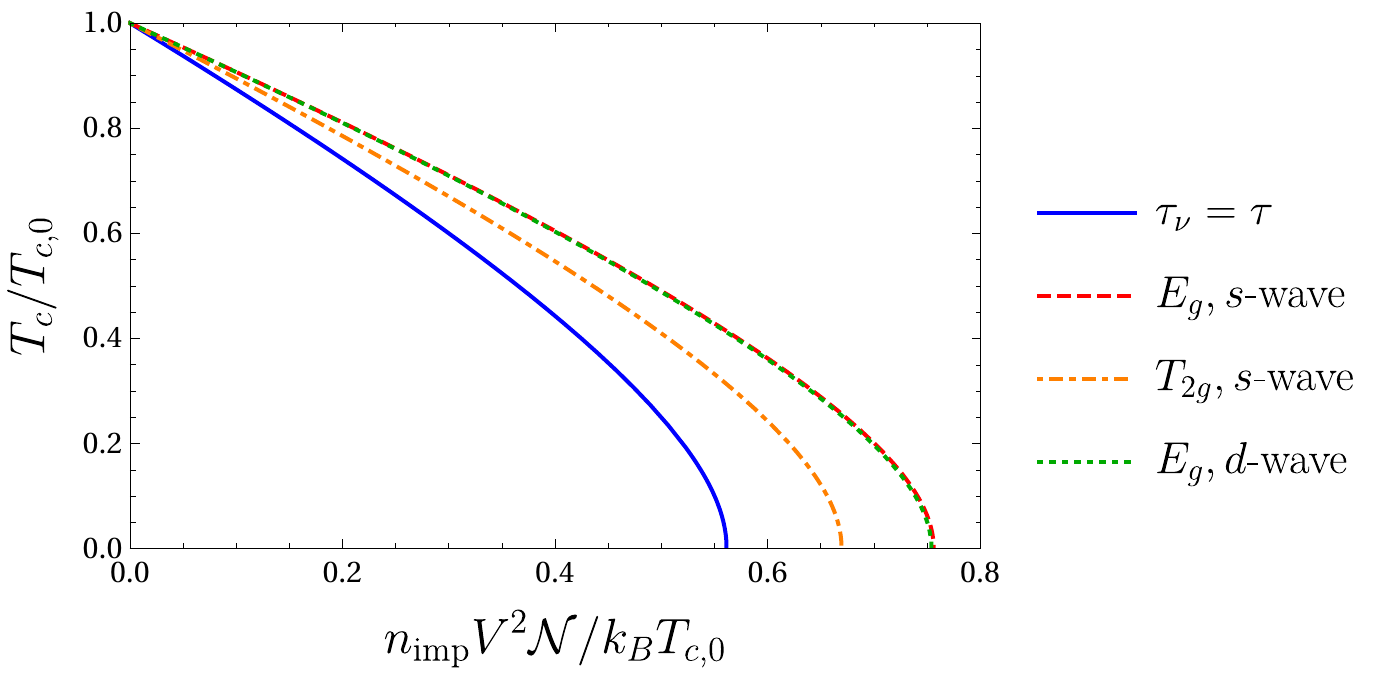}
  \caption{Critical temperature $T_c$ for various gaps in the $E_g$
    and $T_{2g}$ irreps as a function of the disorder 
    strength $n_{\text{imp}}\pi V^2 {\cal N}$
    in YPtBi.  The line $\tau_\nu=\tau$ corresponds to the case where
    the effective SR in~\eq{eq:swaveTc} is equal to the
    normal-state SR, which applies to  pairing states in
    all other  nontrivial irreps.  We
    use parameters for the 
  normal-state Hamiltonian~\eq{eq:LKHam} from Ref.~\cite{Brydon2016}.\label{fig:YPtBi}}
\end{figure}

The six
$s$-wave pairing states in YPtBi are tabulated in
Table~\ref{tab:YPtBi}. Apart from  the $A_{1g}$ singlet state,
there are also five quintet states which pair electrons with total
internal angular momentum $J=2$, and which belong to the
$E_g$ and $T_{2g}$ irreps.
Evaluating~\eq{eq:s-wSig20}, we find that the lowest-order contribution to the anomalous self-energy for
  the $s$-wave gaps  in YPtBi is
\begin{equation}
\Sigma_2^{(0)}= \pi n_{\text{imp}}V^2
\frac{{\cal N}}{4\left|\tilde{\omega}_{n}\right|} \left[ 1+\sum_{l=1}^{5}\lambda_l\langle\hat{\epsilon}_{\bm{k},l}^2\rangle_{\text{FS}}\right]\tilde{\Delta},\label{eq:Sig2_YPtBi}
\end{equation}
where $\lambda_{l}=\pm1$
is tabulated for each channel in Table~\ref{tab:YPtBi}.  Solving~\eq{eq:Sigma2} we obtain the full self-energy
\begin{equation}
\Sigma_2 =
\frac{\bar{\Sigma}_2^{(0)}}{1-\bar{\Sigma}_2^{(0)}}\tilde{\Delta}, \label{eq:Sig2_Sig20}
\end{equation}
where $\Sigma_2^{(0)}=\bar{\Sigma}_2^{(0)}\tilde{\Delta}$. Inserting this into the
linearized gap equation, we find that the critical
temperature $T_{c}$ of the $s$-wave state in channel $\nu$ is given by
the solution of
\begin{equation}
  \log\left(\frac{T_c}{T_{c0}}\right) = \psi\left(\frac{1}{2}\right) -
  \psi\left(\frac{1}{2} + \frac{1}{4\pi k_BT_c\tau_{\nu}}\right) \label{eq:swaveTc}
\end{equation}
where $T_{c0}$ is the critical temperature in the absence of disorder,
$\psi(z)$ is the digamma function, and the effective SR is
\begin{equation}
\frac{1}{\tau_{\nu}} = \frac{1}{2\tau_0}\left(1 - \sum_{l=1}^5\lambda_{l}\langle \hat{\epsilon}_{{\bm k},l}^2\rangle_{\text{FS}} \right)
\end{equation}
with $\tau_0^{-1}= 2\pi n_{\text{imp}}V^2{\cal N}$. We see that
$\lambda_{l}=+1$ decreases the effective SR, whereas
$\lambda_{l}=-1$ brings it closer to the normal-state value $\tau = \tau_0$.
Since all $\lambda_l=1$ for the $A_{1g}$ $s$-wave state, we
find that $\tau_{A_{1g}}^{-1}=0$ and it is hence  insensitive to
disorder, consistent with Anderson's theorem. The effective
SR of the other $s$-wave
states are reduced relative to the normal state value, as in each case there
is one $l$ for which $\lambda_l=+1$. This gives a modest
degree of protection against disorder, as 
shown in~\fig{fig:YPtBi}. 

\begin{table}
  \begin{tabular}{|c||c|c|c|c|c|c|}\hline
     irrep & $A_{1g}$ & \multicolumn{2}{c|}{$E_g$} & \multicolumn{3}{c|}{$T_{2g}$} \\\hline
   $\tilde{\Delta}U_T^\dagger$ & $\mathbb{1}_4$ & $\gamma^1$& $\gamma^2$&
    $\gamma^3$& $\gamma^4$ & $\gamma^5$\\\hline
   nodes & none & line & line & line & line & line\\\hline
    $l$, $\lambda_{l}=1$ & all & 1 & 2 & 3 & 4 & 5\\\hline
    $l$, $\lambda_{l}=-1$ & none & 2,3,4,5 & 1,3,4,5 & 1,2,4,5 & 1,2,3,5 & 1,2,3,4 \\\hline
  \end{tabular}
  \caption{The six $s$-wave pairing states for YPtBi. The first line
    gives the irrep of $O_h$, the second line gives the form of the pairing
    potential in terms of the $\gamma$ matrices defined in the text,
    the third line gives the nodal structure, 
    and the fourth and fifth lines give the
    values of $l$ corresponding to the $\gamma$ matrices for which 
    $\lambda_{l}=1$ and
    $\lambda_{l}=-1$, respectively. \label{tab:YPtBi}}
\end{table}

The enhanced stability of the nontrivial $s$-wave states extends to
other pairing potentials: the critical
temperature for an arbitrary state $\tilde{\Delta}_{\bm k}$ satisfies
\begin{align}
  \log\left(\frac{T_c}{T_{c0}}\right) = &\psi\left(\frac{1}{2}\right) -
  \left(1 - \alpha_\nu(\tilde{\Delta}_{\bm k})\right)\psi\left(\frac{1}{2} + \frac{1}{4\pi
    k_BT_c\tau_0}\right) \notag \\
  & - \alpha_\nu(\tilde{\Delta}_{\bm k})\psi\left(\frac{1}{2} + \frac{1}{4\pi
    k_BT_c\tau_{\nu}}\right)\,, \label{eq:otherTc}
\end{align}
where
\begin{equation}
\alpha_\nu(\tilde{\Delta}_{\bm k}) = \frac{\langle \text{Tr}\{\tilde{\Delta}^\dagger_{\bm k}{\cal
    P}_{\bm k}\tilde{\Delta}_{\nu}{\cal P}_{\bm k}\} \rangle_{\text{FS}}^2}{\langle \text{Tr}\{\tilde{\Delta}^\dagger_{\bm k}{\cal
    P}_{\bm k}\tilde{\Delta}_{\bm k}{\cal P}_{\bm k}\} \rangle_{\text{FS}}\langle \text{Tr}\{\tilde{\Delta}^\dagger_{\nu}{\cal
    P}_{\bm k}\tilde{\Delta}_{\nu}{\cal P}_{\bm k}\} \rangle_{\text{FS}}}\,.
\end{equation}
This parameter measures the similarity of 
$\tilde{\Delta}_{\bm k}$ to the $s$-wave state $\tilde{\Delta}_{\nu}$
at the Fermi surface. The closer $\alpha_\nu$ is to one, the 
more similar these states are to one another, and hence their response
to disorder is also similar. In this way, a general state in an
irrep with a nontrivial $s$-wave pairing potential can also acquire some
robustness against disorder. Indeed, as shown
in~\fig{fig:YPtBi}, the singlet $d$-wave $E_g$  state 
$\tilde{\Delta}_{\bm k}=(\hat{k}_x^2-\hat{k}_y^2)U_T$ 
is almost as stable against disorder as the quintet $s$-wave $E_{g}$
states, reflecting the nearly-identical form of these states at
the Fermi surface. It is instructive to examine the
lowest-order contribution to the anomalous self-energy for this
  state.In particular, the second term inside the brackets of~\eq{eq:s-wSig20} gives the overlap with the 
$s$-wave $E_g$ state $\gamma^1U_T$:
\begin{align}
\Sigma^{(0)}_{2}= -\pi n_{\text{imp}}V^2
\frac{{\cal N}}{2\left|\tilde{\omega}_{n}\right|} \left\langle \hat{\epsilon}_{\bm{k},1}(\hat{k}_x^2-\hat{k}_y^2)\right\rangle_{\text{FS}}\gamma^1U_T,
\end{align}
This is nonzero since $\epsilon_{\bm{k},1} =
\sqrt{3}\beta_1(k_x^2-k_y^2)/2$. The full anomalous self-energy
will have the same form as~\eq{eq:Sig2_Sig20}, where
$\bar{\Sigma}_2^{(0)}$ is the coefficient of $\gamma^1U_T$ in the
expression above.

\section{Application to $\text{Cu$_{x}$Bi$_2$Se$_{3}$}$}\label{sec:CuxBi2Se3}

The low-energy electron states in
Cu$_{x}$Bi$_2$Se$_{3}$ derive from 
$p_z$-like orbitals which are located on opposite sides of each
Bi$_2$Se$_3$ quintuple layer, implying a sublattice degree of freedom. 
The ${\bf k}\cdot{\bf p}$ Hamiltonian for these states to lowest order
in ${\bf k}$ for each term is given by~\cite{Liu2010_topins}
\begin{align}
  H =& -\mu \sigma_0\otimes\eta_0 + m \sigma_0\otimes\eta_x +
  v_zk_z\sigma_0\otimes\eta_y \notag \\
  & +v(k_x\sigma_y - k_y\sigma_x)\otimes\eta_z + \lambda
  k_x(k_x^2-3k_y^2)\sigma_z\otimes\eta_z \label{eq:CuBiSeHam}
\end{align}
where $\sigma_\nu$ and $\eta_\nu$ are the Pauli matrices in spin and
sublattice space, respectively. We choose the $\gamma$ matrices to be
$\vec{\gamma} =
(\sigma_0\otimes\eta_x,\sigma_0\otimes\eta_y,\sigma_x\otimes\eta_z,\sigma_y\otimes\eta_z,\sigma_z\otimes\eta_z)$. 
  The copper intercalation in Cu$_x$Bi$_2$Se$_3$
dopes electrons into the system, giving a Fermi
surface in the upper band.

\begin{table}
  \begin{tabular}{|c||c|c|c|c|c|c|}\hline
     irrep & $A_{1g}$ & $A_{1g}$ & $A_{1u}$ & $A_{2u}$ & \multicolumn{2}{c|}{$E_u$}
     \\\hline
   $\tilde{\Delta}U_T^\dagger$ & $\mathbb{1}_4$ & $\gamma^1$&
     $i\gamma^1\gamma^5$ & $i\gamma^1\gamma^2$ &
    $i\gamma^1\gamma^3$ & $i\gamma^1\gamma^4$ \\\hline
     nodes & none & none & none & point & point & none\\\hline
    $l$, $\lambda_{l}=1$ & all & 1 & 2,3,4 & 3,4,5 & 2,4,5 & 2,3,5  \\\hline
     $l$, $\lambda_{l}=-1$ & none & 2,3,4,5 & 1,5 & 1,2 & 1,3 & 1,4  \\\hline
  \end{tabular}
  \caption{The six $s$-wave pairing states for Cu$_x$Bi$_2$Se$_3$. The first line
    gives the irrep of $D_{3d}$, and the second line gives the form of the pairing
    potential in terms of the $\gamma$ matrices defined in the
    text. The third line gives the nodal structure, while the fourth
    and fifth lines give the 
    values of $l$  corresponding to the $\gamma$ matrices for which  $\lambda_{l}=1$ and
    $\lambda_{l}=-1$, respectively. \label{tab:CuxBi2Se3}}
\end{table}

\begin{figure}
  \includegraphics[width=\columnwidth]{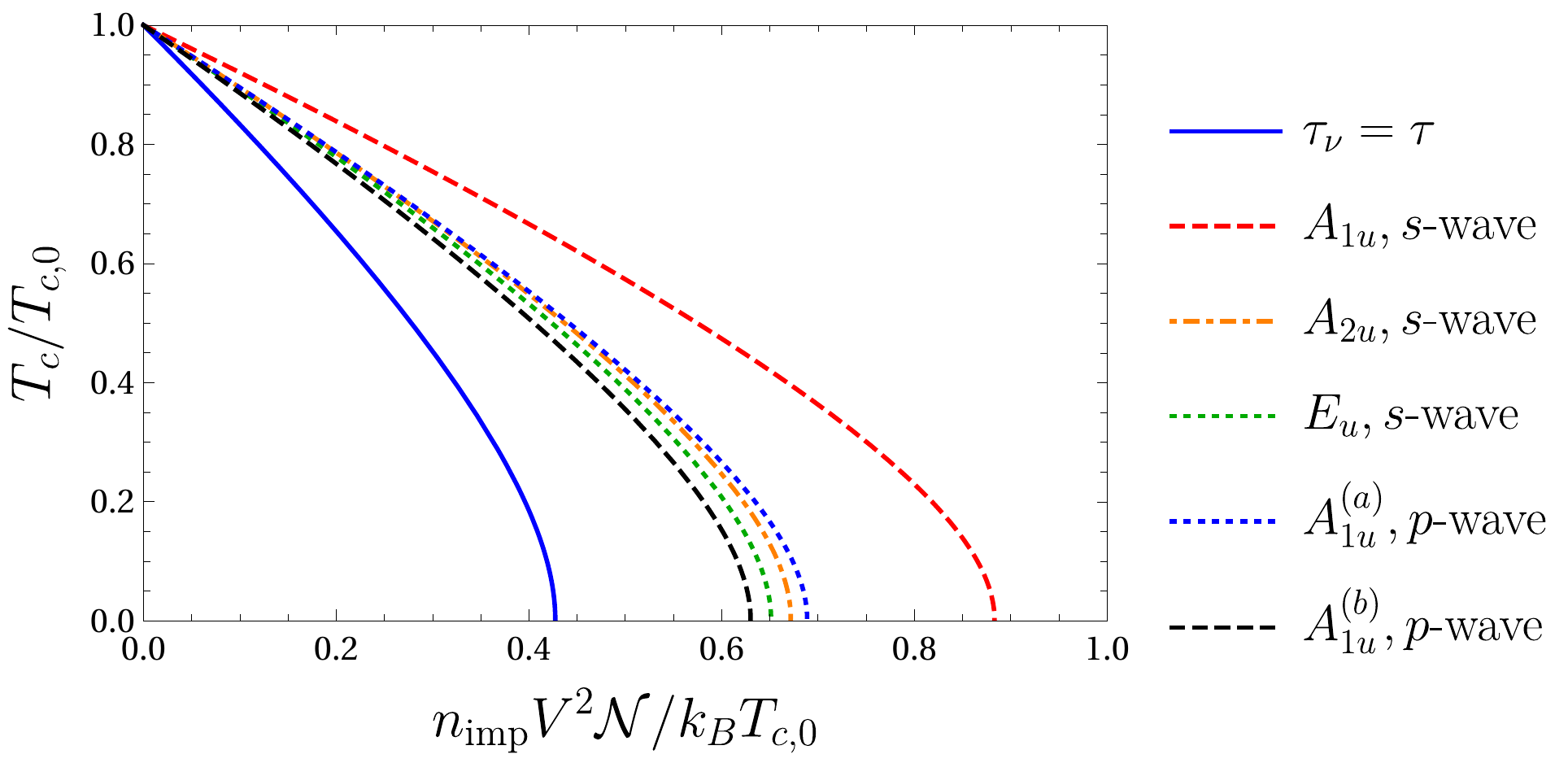}
  \caption{Critical temperature $T_c$ for various gaps in the
    $A_{1u}$, $A_{2u}$, and $E_{u}$ irreps as a function of the disorder 
    strength $n_{\text{imp}}\pi V^2 {\cal N}$
    in Cu$_x$Bi$_2$Se$_3$.  The line $\tau_\nu=\tau$ corresponds to the case where
    the effective SR in~\eq{eq:swaveTc} is equal to the
    normal-state SR, which applies to pairing states in all other nontrivial irreps. We
    use parameters for the 
  normal-state Hamiltonian~\eq{eq:CuBiSeHam} from Ref.~\cite{Liu2010_topins}
  and set $\mu=0.5$eV~\cite{Hashimoto_CuBiSe_2013}. \label{fig:CuBiSe}}
\end{figure}

The six $s$-wave pairing channels in Cu$_x$Bi$_2$Se$_3$ are
summarized in~Table~\ref{tab:CuxBi2Se3}: In addition to two
$A_{1g}$ states, there are four odd-parity
states, which are permitted due to the swapping of the sublattice index
under inversion. The $A_{1g}$ states are insensitive to
disorder~\cite{MichaeliFu2012}, 
although the analysis is more involved than for YPtBi since the
anomalous self-energy includes components from both pairing potentials.
The critical
temperatures of the odd-parity channel $\nu$ is the solution
of~\eq{eq:swaveTc} where  
the effective SR is 
\begin{equation}
\frac{1}{\tau_\nu} = \frac{1}{4\tau_0} \left(1 + 2\langle\hat{\epsilon}_{{\bm k},1}\rangle^2_{\text{FS}} - \sum_{l=1}^5\lambda_l\langle
\hat{\epsilon}_{{\bm k},l}^2\rangle_{\text{FS}}\right)\,, \label{eq:tauCuBiSe}
\end{equation}
and the $\lambda_{l}$ are tabulated in
Table~\ref{tab:CuxBi2Se3}. The second term in the brackets is due to
the nonzero sublattice polarization of the normal-state bands arising from the mass term. 
Our result is consistent with the analysis for the
  $A_{1u}$ channel in~\cite{MichaeliFu2012}.
We plot the
critical temperature as a function of disorder strength for each
channel in~\fig{fig:CuBiSe}.
Other odd-parity
states in Cu$_x$Bi$_2$Se$_3$ also enjoy some degree of protection against 
disorder. In particular, a nontrivial dependence on the sublattice
degrees of freedom is not required. For example, consider the two
$p$-wave spin-triplet sublattice-trivial pairing states in $A_{1u}$:
\begin{align}
A^{(a)}_{1u}: \quad  \tilde{\Delta}_{\bm k} & = \hat{k}_z \sigma_x\otimes\eta_0 \\
A^{(b)}_{1u}: \quad  \tilde{\Delta}_{\bm k}  & = -\hat{k}_x
\sigma_z\otimes\eta_0 + \hat{k}_y  i\sigma_0\otimes\eta_0
\end{align}
As shown in~\fig{fig:CuBiSe}, the robustness of these $p$-wave
states is comparable to the $A_{2u}$ and $E_{u}$ $s$-wave 
states, because of their overlap with the significantly more stable
$A_{1u}$ $s$-wave state.

\section{Discussion}\label{sec:discussion}

Our analysis reveals a remarkable robustness of
the nontrivial $s$-wave pairing states against disorder, which is
manifested by an effective SR which can be greatly reduced from the
normal-state SR. The $s$-wave states play a crucial role, as their
robustness can be shared with, but not exceeded by, any other state in
the same  irrep. 

Since they do not exclusively pair time-reversed partners, the nontrivial
$s$-wave states may involve both pairing of electrons in the same
(intraband pairing) and different (interband pairing) bands. 
Remarkably,
the effective SR of the $s$-wave state in   
channel $\nu$ can be expressed in terms of the Fermi surface average of
the superconducting fitness function $\tilde{F}_{C}(\bm{k})$, which measures the degree of
  interband pairing:
\begin{equation}
\frac{1}{\tau_\nu} =
\frac{1}{\tau} - \frac{1}{\tau_0}\left(1-\tilde{F}_C\right)\,. \label{eq:FAFC}
\end{equation}
Here $\tilde{F}_{C} =\langle \tilde{F}_{C}(\bm{k})\rangle_{FS}=1$ ($0$) implies completely interband (intraband) pairing across the Fermi surface.
The effective SR is reduced, and hence the robustness
against disorder is enhanced, according to the degree that the
$s$-wave state involves intraband pairing.
This result follows from the observation
that $\lambda_l=+1$ ($-1$) when 
$\gamma^l\tilde{\Delta}_{\nu} - \tilde{\Delta}_\nu \gamma^{l,\,\ast}=0$ ($2\gamma^l\tilde{\Delta}_{\nu}$).
We emphasize that~\eq{eq:FAFC} \emph{only} applies to the $s$-wave
states: for other states, the value of $\tilde{F}_C$
does not supply any information about the robustness against
disorder. 

The extreme limit where a nontrivial 
$s$-wave potential
$\tilde{\Delta}_\nu$ is perfectly fit
(i.e. $\tilde{F}_C=0$) is instructive. As shown in~\cite{FuBerg2010}, the
Bogoliubov-de Gennes Hamiltonian can then be mapped to
that for the trivial $s$-wave state using 
$c_{\bm k} \rightarrow \exp(i\frac{\pi}{4}\tilde{\Delta}_\nu U_T^\dagger)c_{\bm
  k}$. This global transformation leaves the impurity Hamiltonian~\eq{eq:impHam}
invariant, and so  the nontrivial $s$-wave
state is insensitive to nonmagnetic disorder, giving a generalization
of Anderson's theorem~\cite{Scheurer2016}.  
The Hamiltonian will generally contain terms which violate the
  fitness condition (i.e. $\tilde{F}_C>0$), however,
which spoils this correspondence. Nevertheless, the nontrivial
$s$-wave state will retain some robustness 
against disorder.

This effect is very sensitive to the material
parameters. For example, it is known that the
robustness of the  odd-parity
$s$-wave states in Cu$_x$Bi$_2$Se$_3$ is enhanced
by reducing the mass term $m$
in~\eq{eq:CuBiSeHam}~\cite{MichaeliFu2012,Nagai2015}. This is immediately evident in our framework, where the effective scattering rate is always enhanced by a finite mass. For the odd-parity s-wave gaps, $\tilde{\Delta}=i\gamma^1\gamma^jU_T$, the enhancement is $\tau_0/\tau_{\nu}= \hat{m}^2+\langle\hat{\epsilon}_{{\bm k},j}^2\rangle_{\text{FS}}/2$, and the $A_{1u}$ state is the most stable as $\langle\hat{\epsilon}_{{\bm k},5}^2\rangle_{\text{FS}}$ is the smallest component of the Hamiltonian. This is a direct consequence of the fact that the mass term in the Hamiltonian is proportional to the nontrivial inversion symmetry operator ${\cal I}=\gamma^1$, and thus the odd parity gaps must by definition have $\lambda_1=-1$.   

Equation~(\ref{eq:FAFC})
gives a simple diagnostic for the existence of a highly-robust
nontrivial irrep in a general system: there must be an $s$-wave state
in this irrep such that $\tilde{F}_C\ll 1$. A nontrivial inversion
operator is highly desirable: 
in this case, the odd-parity $s$-wave states involve the product of
two $\gamma$ matrices (one of which is the inversion operator), and
hence  commute with three $\gamma$ matrices in the  general
Hamiltonian~\eq{eq:genH}. In contrast, the
even-parity $s$-wave states commute with 
only one $\gamma$ matrix in the Hamiltonian when inversion is
trivial. Assuming roughly equal  
values of all the coefficients $\epsilon_{{\bm k},l}$ at the Fermi
surface, $\tilde{F}_C$ will typically be
smaller for the $s$-wave states in the system with nontrivial
inversion. This is exemplified by the greater robustness of the
$s$-wave states in Cu$_x$Bi$_2$Se$_3$ compared to YPtBi.

The analysis presented above has focused entirely on
  understanding the role of the spin-orbital
  texture of the normal-state bands. The impurity physics of
  superconductors is a rich field~\cite{RMP_impurity}, and although
  the self-consistent Born approximation utilized here can
  successfully account for the pair-breaking physics in the dilute
  impurity limit, effects beyond this approximation can be
  important. For example, it has recently been shown that the enhancement of
  the local density of states due to the presence of 
  resonant levels at the impurity sites can increase the critical
  temperature in unconventional multiorbital superconductors above the
  clean-limit result~\cite{Gastiasoro2018}.  We nevertheless expect
  our results to remain qualitatively valid for more sophisticated
  treatments, as the spin-orbital
  texture and the superconducting fitness are properties of the
  clean-limit Bogoliubov-de Gennes equations. Indeed, the
  role of the mass term in controlling the robustness against disorder
  in Cu$_x$Bi$_2$Se$_3$ has been numerically confirmed using a
  self-consistent $T$-matrix theory~\cite{Nagai2015}. Extending our
  theory beyond the self-consistent Born approximation is a promising
  direction for future work.

During final preparation of our manuscript we became aware of a
similar analysis in Ref.~\cite{Ramires2019}. However, our results
for the effective SR disagree: whereas we find that this
involves the superconducting fitness with respect to the normal-state 
Hamiltonian, in Ref.~\cite{Ramires2019} the 
superconducting fitness with respect to the impurity
Hamiltonian appears. This gives a complete insensitivity of the
pairing state to
disorder, in contrast to the parametric enhancement of the robustness
found here, and disagrees with previous
studies~\cite{MichaeliFu2012,Scheurer2016,Nagai2015}.  

\section{Conclusions}

In this manuscript we have shown that
unconventional superconducting states in multiband systems are
generically  less sensitive to the presence of nonmagnetic
disorder than unconventional states in single-band materials,
  due to the spin-orbital texture of the normal-state bands. The
enhanced stability occurs for pairing states in irreps for
which there  is a
nontrivial $s$-wave state. The degree to which an $s$-wave state is
robust against disorder can be quantified in terms of the Fermi
surface average of the superconducting fitness parameter, and provides
an upper bound for the stability of all other 
states in the same irrep. Our theory offers a straightforward way to
assess the robustness against disorder of unconventional pairing states for any
multiband system, and can thus guide the search for novel
superconducting states.

\begin{acknowledgments}
  
This work was supported by the Marsden Fund Council from Government
funding, managed by Royal Society Te Ap\={a}rangi.


\end{acknowledgments}

\bibliography{references}

\begin{thebibliography}{32}%
\makeatletter
\providecommand \@ifxundefined [1]{%
 \@ifx{#1\undefined}
}%
\providecommand \@ifnum [1]{%
 \ifnum #1\expandafter \@firstoftwo
 \else \expandafter \@secondoftwo
 \fi
}%
\providecommand \@ifx [1]{%
 \ifx #1\expandafter \@firstoftwo
 \else \expandafter \@secondoftwo
 \fi
}%
\providecommand \natexlab [1]{#1}%
\providecommand \enquote  [1]{``#1''}%
\providecommand \bibnamefont  [1]{#1}%
\providecommand \bibfnamefont [1]{#1}%
\providecommand \citenamefont [1]{#1}%
\providecommand \href@noop [0]{\@secondoftwo}%
\providecommand \href [0]{\begingroup \@sanitize@url \@href}%
\providecommand \@href[1]{\@@startlink{#1}\@@href}%
\providecommand \@@href[1]{\endgroup#1\@@endlink}%
\providecommand \@sanitize@url [0]{\catcode `\\12\catcode `\$12\catcode
  `\&12\catcode `\#12\catcode `\^12\catcode `\_12\catcode `\%12\relax}%
\providecommand \@@startlink[1]{}%
\providecommand \@@endlink[0]{}%
\providecommand \url  [0]{\begingroup\@sanitize@url \@url }%
\providecommand \@url [1]{\endgroup\@href {#1}{\urlprefix }}%
\providecommand \urlprefix  [0]{URL }%
\providecommand \Eprint [0]{\href }%
\providecommand \doibase [0]{http://dx.doi.org/}%
\providecommand \selectlanguage [0]{\@gobble}%
\providecommand \bibinfo  [0]{\@secondoftwo}%
\providecommand \bibfield  [0]{\@secondoftwo}%
\providecommand \translation [1]{[#1]}%
\providecommand \BibitemOpen [0]{}%
\providecommand \bibitemStop [0]{}%
\providecommand \bibitemNoStop [0]{.\EOS\space}%
\providecommand \EOS [0]{\spacefactor3000\relax}%
\providecommand \BibitemShut  [1]{\csname bibitem#1\endcsname}%
\let\auto@bib@innerbib\@empty
\bibitem [{\citenamefont {Mineev}\ and\ \citenamefont
  {Samokhin}(1999)}]{Mineev1999}%
  \BibitemOpen
  \bibfield  {author} {\bibinfo {author} {\bibfnamefont {V.~P.}\ \bibnamefont
  {Mineev}}\ and\ \bibinfo {author} {\bibfnamefont {K.~V.}\ \bibnamefont
  {Samokhin}},\ }\href@noop {} {\emph {\bibinfo {title} {{Introduction to
  Unconventional Superconductivity}}}}\ (\bibinfo  {publisher} {Gordon and
  Breach Science Publishers},\ \bibinfo {year} {1999})\BibitemShut {NoStop}%
\bibitem [{\citenamefont {Anderson}(1959)}]{Anderson1959}%
  \BibitemOpen
  \bibfield  {author} {\bibinfo {author} {\bibfnamefont {P.~W.}\ \bibnamefont
  {Anderson}},\ }\bibfield  {title} {\enquote {\bibinfo {title} {{Theory of
  dirty superconductors}},}\ }\href {\doibase 10.1016/0022-3697(59)90036-8}
  {\bibfield  {journal} {\bibinfo  {journal} {J. Phys. Chem. Solids}\ }\textbf
  {\bibinfo {volume} {11}},\ \bibinfo {pages} {26–30} (\bibinfo {year}
  {1959})}\BibitemShut {NoStop}%
\bibitem [{\citenamefont {Ong}\ \emph {et~al.}(2016)\citenamefont {Ong},
  \citenamefont {Coleman},\ and\ \citenamefont {Schmalian}}]{Ong2016}%
  \BibitemOpen
  \bibfield  {author} {\bibinfo {author} {\bibfnamefont {T.}~\bibnamefont
  {Ong}}, \bibinfo {author} {\bibfnamefont {P.}~\bibnamefont {Coleman}}, \ and\
  \bibinfo {author} {\bibfnamefont {J.}~\bibnamefont {Schmalian}},\ }\bibfield
  {title} {\enquote {\bibinfo {title} {Concealed $d$-wave pairs in the $s_\pm$
  condensate of iron-based superconductors},}\ }\href {\doibase
  10.1073/pnas.1523064113} {\bibfield  {journal} {\bibinfo  {journal}
  {Proceedings of the National Academy of Sciences}\ }\textbf {\bibinfo
  {volume} {113}},\ \bibinfo {pages} {5486--5491} (\bibinfo {year}
  {2016})}\BibitemShut {NoStop}%
\bibitem [{\citenamefont {Vafek}\ and\ \citenamefont
  {Chubukov}(2017)}]{Vafek2016}%
  \BibitemOpen
  \bibfield  {author} {\bibinfo {author} {\bibfnamefont {O.}~\bibnamefont
  {Vafek}}\ and\ \bibinfo {author} {\bibfnamefont {A.~V.}\ \bibnamefont
  {Chubukov}},\ }\bibfield  {title} {\enquote {\bibinfo {title} {{Hund
  Interaction, Spin-Orbit Coupling, and the Mechanism of Superconductivity in
  Strongly Hole-Doped Iron Pnictides}},}\ }\href {\doibase
  10.1103/PhysRevLett.118.087003} {\bibfield  {journal} {\bibinfo  {journal}
  {Phys. Rev. Lett.}\ }\textbf {\bibinfo {volume} {118}},\ \bibinfo {pages}
  {087003} (\bibinfo {year} {2017})}\BibitemShut {NoStop}%
\bibitem [{\citenamefont {Agterberg}\ \emph {et~al.}(2017)\citenamefont
  {Agterberg}, \citenamefont {Shishidou}, \citenamefont {O'Halloran},
  \citenamefont {Brydon},\ and\ \citenamefont {Weinert}}]{AgterbergFeSe2017}%
  \BibitemOpen
  \bibfield  {author} {\bibinfo {author} {\bibfnamefont {D.~F.}\ \bibnamefont
  {Agterberg}}, \bibinfo {author} {\bibfnamefont {T.}~\bibnamefont
  {Shishidou}}, \bibinfo {author} {\bibfnamefont {J.}~\bibnamefont
  {O'Halloran}}, \bibinfo {author} {\bibfnamefont {P.~M.~R.}\ \bibnamefont
  {Brydon}}, \ and\ \bibinfo {author} {\bibfnamefont {M.}~\bibnamefont
  {Weinert}},\ }\bibfield  {title} {\enquote {\bibinfo {title} {{Resilient
  Nodeless $d$-Wave Superconductivity in Monolayer FeSe}},}\ }\href {\doibase
  10.1103/PhysRevLett.119.267001} {\bibfield  {journal} {\bibinfo  {journal}
  {Phys. Rev. Lett.}\ }\textbf {\bibinfo {volume} {119}},\ \bibinfo {pages}
  {267001} (\bibinfo {year} {2017})}\BibitemShut {NoStop}%
\bibitem [{\citenamefont {Kawakami}\ \emph {et~al.}(2018)\citenamefont
  {Kawakami}, \citenamefont {Okamura}, \citenamefont {Kobayashi},\ and\
  \citenamefont {Sato}}]{Kawakami2018}%
  \BibitemOpen
  \bibfield  {author} {\bibinfo {author} {\bibfnamefont {T.}~\bibnamefont
  {Kawakami}}, \bibinfo {author} {\bibfnamefont {T.}~\bibnamefont {Okamura}},
  \bibinfo {author} {\bibfnamefont {S.}~\bibnamefont {Kobayashi}}, \ and\
  \bibinfo {author} {\bibfnamefont {M.}~\bibnamefont {Sato}},\ }\bibfield
  {title} {\enquote {\bibinfo {title} {{Topological Crystalline Materials of
  $J=3/2$ Electrons: Antiperovskites, Dirac Points, and High Winding
  Topological Superconductivity}},}\ }\href {\doibase
  10.1103/PhysRevX.8.041026} {\bibfield  {journal} {\bibinfo  {journal} {Phys.
  Rev. X}\ }\textbf {\bibinfo {volume} {8}},\ \bibinfo {pages} {041026}
  (\bibinfo {year} {2018})}\BibitemShut {NoStop}%
\bibitem [{\citenamefont {Oiwa}\ \emph {et~al.}(2018)\citenamefont {Oiwa},
  \citenamefont {Yanagi},\ and\ \citenamefont {Kusunose}}]{Oiwa2018}%
  \BibitemOpen
  \bibfield  {author} {\bibinfo {author} {\bibfnamefont {R.}~\bibnamefont
  {Oiwa}}, \bibinfo {author} {\bibfnamefont {Y.}~\bibnamefont {Yanagi}}, \ and\
  \bibinfo {author} {\bibfnamefont {H.}~\bibnamefont {Kusunose}},\ }\bibfield
  {title} {\enquote {\bibinfo {title} {{Theory of superconductivity in
  hole-doped monolayer ${\mathrm{MoS}}_{2}$}},}\ }\href {\doibase
  10.1103/PhysRevB.98.064509} {\bibfield  {journal} {\bibinfo  {journal} {Phys.
  Rev. B}\ }\textbf {\bibinfo {volume} {98}},\ \bibinfo {pages} {064509}
  (\bibinfo {year} {2018})}\BibitemShut {NoStop}%
\bibitem [{\citenamefont {M\"ockli}\ and\ \citenamefont
  {Khodas}(2018)}]{Moeckli2018}%
  \BibitemOpen
  \bibfield  {author} {\bibinfo {author} {\bibfnamefont {D.}~\bibnamefont
  {M\"ockli}}\ and\ \bibinfo {author} {\bibfnamefont {M.}~\bibnamefont
  {Khodas}},\ }\bibfield  {title} {\enquote {\bibinfo {title} {Robust
  parity-mixed superconductivity in disordered monolayer transition metal
  dichalcogenides},}\ }\href {\doibase 10.1103/PhysRevB.98.144518} {\bibfield
  {journal} {\bibinfo  {journal} {Phys. Rev. B}\ }\textbf {\bibinfo {volume}
  {98}},\ \bibinfo {pages} {144518} (\bibinfo {year} {2018})}\BibitemShut
  {NoStop}%
\bibitem [{\citenamefont {Fu}\ and\ \citenamefont {Berg}(2010)}]{FuBerg2010}%
  \BibitemOpen
  \bibfield  {author} {\bibinfo {author} {\bibfnamefont {L.}~\bibnamefont
  {Fu}}\ and\ \bibinfo {author} {\bibfnamefont {E.}~\bibnamefont {Berg}},\
  }\bibfield  {title} {\enquote {\bibinfo {title} {{Odd-Parity Topological
  Superconductors: Theory and Application to
  ${\mathrm{Cu}}_{x}{\mathrm{Bi}}_{2}{\mathrm{Se}}_{3}$}},}\ }\href {\doibase
  10.1103/PhysRevLett.105.097001} {\bibfield  {journal} {\bibinfo  {journal}
  {Phys. Rev. Lett.}\ }\textbf {\bibinfo {volume} {105}},\ \bibinfo {pages}
  {097001} (\bibinfo {year} {2010})}\BibitemShut {NoStop}%
\bibitem [{\citenamefont {Brydon}\ \emph {et~al.}(2016)\citenamefont {Brydon},
  \citenamefont {Wang}, \citenamefont {Weinert},\ and\ \citenamefont
  {Agterberg}}]{Brydon2016}%
  \BibitemOpen
  \bibfield  {author} {\bibinfo {author} {\bibfnamefont {P.~M.~R.}\
  \bibnamefont {Brydon}}, \bibinfo {author} {\bibfnamefont {L.}~\bibnamefont
  {Wang}}, \bibinfo {author} {\bibfnamefont {M.}~\bibnamefont {Weinert}}, \
  and\ \bibinfo {author} {\bibfnamefont {D.~F.}\ \bibnamefont {Agterberg}},\
  }\bibfield  {title} {\enquote {\bibinfo {title} {{Pairing of $j=3/2$ Fermions
  in Half-{H}eusler Superconductors}},}\ }\href {\doibase
  10.1103/PhysRevLett.116.177001} {\bibfield  {journal} {\bibinfo  {journal}
  {Phys. Rev. Lett.}\ }\textbf {\bibinfo {volume} {116}},\ \bibinfo {pages}
  {177001} (\bibinfo {year} {2016})}\BibitemShut {NoStop}%
\bibitem [{\citenamefont {Matano}\ \emph {et~al.}(2016)\citenamefont {Matano},
  \citenamefont {Kriener}, \citenamefont {Segawa}, \citenamefont {Ando},\ and\
  \citenamefont {Zheng}}]{Matano2016}%
  \BibitemOpen
  \bibfield  {author} {\bibinfo {author} {\bibfnamefont {K.}~\bibnamefont
  {Matano}}, \bibinfo {author} {\bibfnamefont {M.}~\bibnamefont {Kriener}},
  \bibinfo {author} {\bibfnamefont {K.}~\bibnamefont {Segawa}}, \bibinfo
  {author} {\bibfnamefont {Y.}~\bibnamefont {Ando}}, \ and\ \bibinfo {author}
  {\bibfnamefont {G.-q.}\ \bibnamefont {Zheng}},\ }\bibfield  {title} {\enquote
  {\bibinfo {title} {{Spin-rotation symmetry breaking in the superconducting
  state of Cu$_x$Bi$_2$Se$_3$}},}\ }\href
  {https://www.nature.com/articles/nphys3781\#supplementary-information}
  {\bibfield  {journal} {\bibinfo  {journal} {Nature Physics}\ }\textbf
  {\bibinfo {volume} {12}},\ \bibinfo {pages} {852} (\bibinfo {year}
  {2016})}\BibitemShut {NoStop}%
\bibitem [{\citenamefont {Yonezawa}\ \emph {et~al.}(2016)\citenamefont
  {Yonezawa}, \citenamefont {Tajiri}, \citenamefont {Nakata}, \citenamefont
  {Nagai}, \citenamefont {Wang}, \citenamefont {Segawa}, \citenamefont {Ando},\
  and\ \citenamefont {Maeno}}]{Yonezawa2016}%
  \BibitemOpen
  \bibfield  {author} {\bibinfo {author} {\bibfnamefont {S.}~\bibnamefont
  {Yonezawa}}, \bibinfo {author} {\bibfnamefont {K.}~\bibnamefont {Tajiri}},
  \bibinfo {author} {\bibfnamefont {S.}~\bibnamefont {Nakata}}, \bibinfo
  {author} {\bibfnamefont {Y.}~\bibnamefont {Nagai}}, \bibinfo {author}
  {\bibfnamefont {Z.}~\bibnamefont {Wang}}, \bibinfo {author} {\bibfnamefont
  {K.}~\bibnamefont {Segawa}}, \bibinfo {author} {\bibfnamefont
  {Y.}~\bibnamefont {Ando}}, \ and\ \bibinfo {author} {\bibfnamefont
  {Y.}~\bibnamefont {Maeno}},\ }\bibfield  {title} {\enquote {\bibinfo {title}
  {{Thermodynamic evidence for nematic superconductivity in
  {Cu$_x$Bi$_2$Se$_3$}}},}\ }\href {http://dx.doi.org/10.1038/nphys3907}
  {\bibfield  {journal} {\bibinfo  {journal} {Nature Physics}\ }\textbf
  {\bibinfo {volume} {13}},\ \bibinfo {pages} {123} (\bibinfo {year}
  {2016})}\BibitemShut {NoStop}%
\bibitem [{\citenamefont {Tao}\ \emph {et~al.}(2018)\citenamefont {Tao},
  \citenamefont {Yan}, \citenamefont {Liu}, \citenamefont {Wang}, \citenamefont
  {Ando}, \citenamefont {Wang}, \citenamefont {Zhang},\ and\ \citenamefont
  {Feng}}]{Tao2018}%
  \BibitemOpen
  \bibfield  {author} {\bibinfo {author} {\bibfnamefont {R.}~\bibnamefont
  {Tao}}, \bibinfo {author} {\bibfnamefont {Y.-J.}\ \bibnamefont {Yan}},
  \bibinfo {author} {\bibfnamefont {X.}~\bibnamefont {Liu}}, \bibinfo {author}
  {\bibfnamefont {Z.-W.}\ \bibnamefont {Wang}}, \bibinfo {author}
  {\bibfnamefont {Y.}~\bibnamefont {Ando}}, \bibinfo {author} {\bibfnamefont
  {Q.-H.}\ \bibnamefont {Wang}}, \bibinfo {author} {\bibfnamefont
  {T.}~\bibnamefont {Zhang}}, \ and\ \bibinfo {author} {\bibfnamefont {D.-L.}\
  \bibnamefont {Feng}},\ }\bibfield  {title} {\enquote {\bibinfo {title}
  {{Direct Visualization of the Nematic Superconductivity in
  ${\mathrm{Cu}}_{x}{\mathrm{Bi}}_{2}{\mathrm{Se}}_{3}$}},}\ }\href {\doibase
  10.1103/PhysRevX.8.041024} {\bibfield  {journal} {\bibinfo  {journal} {Phys.
  Rev. X}\ }\textbf {\bibinfo {volume} {8}},\ \bibinfo {pages} {041024}
  (\bibinfo {year} {2018})}\BibitemShut {NoStop}%
\bibitem [{\citenamefont {Fu}(2014)}]{Fu2014}%
  \BibitemOpen
  \bibfield  {author} {\bibinfo {author} {\bibfnamefont {L.}~\bibnamefont
  {Fu}},\ }\bibfield  {title} {\enquote {\bibinfo {title} {Odd-parity
  topological superconductor with nematic order: Application to
  {Cu}$_{x}${Bi}$_{2}${Se}$_{3}$},}\ }\href {\doibase
  10.1103/PhysRevB.90.100509} {\bibfield  {journal} {\bibinfo  {journal} {Phys.
  Rev. B}\ }\textbf {\bibinfo {volume} {90}},\ \bibinfo {pages} {100509(R)}
  (\bibinfo {year} {2014})}\BibitemShut {NoStop}%
\bibitem [{\citenamefont {Kim}\ \emph {et~al.}(2018)\citenamefont {Kim},
  \citenamefont {Wang}, \citenamefont {Nakajima}, \citenamefont {Hu},
  \citenamefont {Ziemak}, \citenamefont {Syers}, \citenamefont {Wang},
  \citenamefont {Hodovanets}, \citenamefont {Denlinger}, \citenamefont
  {Brydon}, \citenamefont {Agterberg}, \citenamefont {Tanatar}, \citenamefont
  {Prozorov},\ and\ \citenamefont {Paglione}}]{Kim2018}%
  \BibitemOpen
  \bibfield  {author} {\bibinfo {author} {\bibfnamefont {H.}~\bibnamefont
  {Kim}}, \bibinfo {author} {\bibfnamefont {K.}~\bibnamefont {Wang}}, \bibinfo
  {author} {\bibfnamefont {Y.}~\bibnamefont {Nakajima}}, \bibinfo {author}
  {\bibfnamefont {R.}~\bibnamefont {Hu}}, \bibinfo {author} {\bibfnamefont
  {S.}~\bibnamefont {Ziemak}}, \bibinfo {author} {\bibfnamefont
  {P.}~\bibnamefont {Syers}}, \bibinfo {author} {\bibfnamefont
  {L.}~\bibnamefont {Wang}}, \bibinfo {author} {\bibfnamefont {H.}~\bibnamefont
  {Hodovanets}}, \bibinfo {author} {\bibfnamefont {J.~D.}\ \bibnamefont
  {Denlinger}}, \bibinfo {author} {\bibfnamefont {P.~M.~R.}\ \bibnamefont
  {Brydon}}, \bibinfo {author} {\bibfnamefont {D.~F.}\ \bibnamefont
  {Agterberg}}, \bibinfo {author} {\bibfnamefont {M.~A.}\ \bibnamefont
  {Tanatar}}, \bibinfo {author} {\bibfnamefont {R.}~\bibnamefont {Prozorov}}, \
  and\ \bibinfo {author} {\bibfnamefont {J.}~\bibnamefont {Paglione}},\
  }\bibfield  {title} {\enquote {\bibinfo {title} {{Beyond Triplet:
  Unconventional Superconductivity in a Spin-$3/2$ Topological Semimetal}},}\
  }\href {\doibase 10.1126/sciadv.aao4513} {\bibfield  {journal} {\bibinfo
  {journal} {Sci. Adv.}\ }\textbf {\bibinfo {volume} {4}},\ \bibinfo {pages}
  {eaao4513} (\bibinfo {year} {2018})}\BibitemShut {NoStop}%
\bibitem [{\citenamefont {Timm}\ \emph {et~al.}(2017)\citenamefont {Timm},
  \citenamefont {Schnyder}, \citenamefont {Agterberg},\ and\ \citenamefont
  {Brydon}}]{Timm2017}%
  \BibitemOpen
  \bibfield  {author} {\bibinfo {author} {\bibfnamefont {C.}~\bibnamefont
  {Timm}}, \bibinfo {author} {\bibfnamefont {A.~P.}\ \bibnamefont {Schnyder}},
  \bibinfo {author} {\bibfnamefont {D.~F.}\ \bibnamefont {Agterberg}}, \ and\
  \bibinfo {author} {\bibfnamefont {P.~M.~R.}\ \bibnamefont {Brydon}},\
  }\bibfield  {title} {\enquote {\bibinfo {title} {{Inflated nodes and surface
  states in superconducting half-{H}eusler compounds}},}\ }\href {\doibase
  10.1103/PhysRevB.96.094526} {\bibfield  {journal} {\bibinfo  {journal} {Phys.
  Rev. B}\ }\textbf {\bibinfo {volume} {96}},\ \bibinfo {pages} {094526}
  (\bibinfo {year} {2017})}\BibitemShut {NoStop}%
\bibitem [{\citenamefont {Yip}(2013)}]{Yip2013}%
  \BibitemOpen
  \bibfield  {author} {\bibinfo {author} {\bibfnamefont {S.-K.}\ \bibnamefont
  {Yip}},\ }\bibfield  {title} {\enquote {\bibinfo {title} {{Models of
  superconducting Cu:Bi${}_{2}$Se${}_{3}$: Single- versus two-band
  description}},}\ }\href {\doibase 10.1103/PhysRevB.87.104505} {\bibfield
  {journal} {\bibinfo  {journal} {Phys. Rev. B}\ }\textbf {\bibinfo {volume}
  {87}},\ \bibinfo {pages} {104505} (\bibinfo {year} {2013})}\BibitemShut
  {NoStop}%
\bibitem [{\citenamefont {Fomin}(2018)}]{Fomin2018}%
  \BibitemOpen
  \bibfield  {author} {\bibinfo {author} {\bibfnamefont {I.~A.}\ \bibnamefont
  {Fomin}},\ }\bibfield  {title} {\enquote {\bibinfo {title} {{Analog of the
  Anderson Theorem for the Polar Phase of Liquid $^3$He in a Nematic
  Aerogel}},}\ }\href {\doibase 10.1134/S106377611811002X} {\bibfield
  {journal} {\bibinfo  {journal} {J. Exp. Theor. Phys}\ }\textbf {\bibinfo
  {volume} {127}},\ \bibinfo {pages} {933} (\bibinfo {year}
  {2018})}\BibitemShut {NoStop}%
\bibitem [{\citenamefont {Eltsov}\ \emph {et~al.}(2019)\citenamefont {Eltsov},
  \citenamefont {Kamppinen}, \citenamefont {Rysti},\ and\ \citenamefont
  {Volovik}}]{Eltsov2019}%
  \BibitemOpen
  \bibfield  {author} {\bibinfo {author} {\bibfnamefont {V.~B.}\ \bibnamefont
  {Eltsov}}, \bibinfo {author} {\bibfnamefont {T.}~\bibnamefont {Kamppinen}},
  \bibinfo {author} {\bibfnamefont {J.}~\bibnamefont {Rysti}}, \ and\ \bibinfo
  {author} {\bibfnamefont {G.~E.}\ \bibnamefont {Volovik}},\ }\href
  {https://arxiv.org/abs/1908.01645} {\enquote {\bibinfo {title} {{Topological
  nodal line in superfluid $^3$He and the Anderson theorem}},}\ } (\bibinfo
  {year} {2019}),\ \Eprint {http://arxiv.org/abs/1908.01645} {arXiv:1908.01645}
  \BibitemShut {NoStop}%
\bibitem [{\citenamefont {Michaeli}\ and\ \citenamefont
  {Fu}(2012)}]{MichaeliFu2012}%
  \BibitemOpen
  \bibfield  {author} {\bibinfo {author} {\bibfnamefont {K.}~\bibnamefont
  {Michaeli}}\ and\ \bibinfo {author} {\bibfnamefont {L.}~\bibnamefont {Fu}},\
  }\bibfield  {title} {\enquote {\bibinfo {title} {{Spin-Orbit Locking as a
  Protection Mechanism of the Odd-Parity Superconducting State against
  Disorder}},}\ }\href {\doibase 10.1103/PhysRevLett.109.187003} {\bibfield
  {journal} {\bibinfo  {journal} {Phys. Rev. Lett.}\ }\textbf {\bibinfo
  {volume} {109}},\ \bibinfo {pages} {187003} (\bibinfo {year}
  {2012})}\BibitemShut {NoStop}%
\bibitem [{\citenamefont {Nagai}(2015)}]{Nagai2015}%
  \BibitemOpen
  \bibfield  {author} {\bibinfo {author} {\bibfnamefont {Y.}~\bibnamefont
  {Nagai}},\ }\bibfield  {title} {\enquote {\bibinfo {title} {{Robust
  superconductivity with nodes in the superconducting topological insulator
  ${\text{Cu}}_{x}{\text{Bi}}_{2}{\text{Se}}_{3}$: Zeeman orbital field and
  nonmagnetic impurities}},}\ }\href {\doibase 10.1103/PhysRevB.91.060502}
  {\bibfield  {journal} {\bibinfo  {journal} {Phys. Rev. B}\ }\textbf {\bibinfo
  {volume} {91}},\ \bibinfo {pages} {060502(R)} (\bibinfo {year}
  {2015})}\BibitemShut {NoStop}%
\bibitem [{\citenamefont {Scheurer}(2016)}]{Scheurer2016}%
  \BibitemOpen
  \bibfield  {author} {\bibinfo {author} {\bibfnamefont {M.~S.}\ \bibnamefont
  {Scheurer}},\ }\emph {\bibinfo {title} {{Mechanism, symmetry and topology of
  ordered phases in correlated systems}}},\ \href {\doibase
  10.5445/IR/1000056491} {Ph.D. thesis} (\bibinfo {year} {2016})\BibitemShut
  {NoStop}%
\bibitem [{\citenamefont {Ramires}\ and\ \citenamefont
  {Sigrist}(2016)}]{Ramires2016}%
  \BibitemOpen
  \bibfield  {author} {\bibinfo {author} {\bibfnamefont {A.}~\bibnamefont
  {Ramires}}\ and\ \bibinfo {author} {\bibfnamefont {M.}~\bibnamefont
  {Sigrist}},\ }\bibfield  {title} {\enquote {\bibinfo {title} {{Identifying
  detrimental effects for multiorbital superconductivity: Application to
  ${\mathrm{Sr}}_{2}{\mathrm{RuO}}_{4}$}},}\ }\href {\doibase
  10.1103/PhysRevB.94.104501} {\bibfield  {journal} {\bibinfo  {journal} {Phys.
  Rev. B}\ }\textbf {\bibinfo {volume} {94}},\ \bibinfo {pages} {104501}
  (\bibinfo {year} {2016})}\BibitemShut {NoStop}%
\bibitem [{\citenamefont {Ramires}\ \emph {et~al.}(2018)\citenamefont
  {Ramires}, \citenamefont {Agterberg},\ and\ \citenamefont
  {Sigrist}}]{Ramires2018}%
  \BibitemOpen
  \bibfield  {author} {\bibinfo {author} {\bibfnamefont {A.}~\bibnamefont
  {Ramires}}, \bibinfo {author} {\bibfnamefont {D.~F.}\ \bibnamefont
  {Agterberg}}, \ and\ \bibinfo {author} {\bibfnamefont {M.}~\bibnamefont
  {Sigrist}},\ }\bibfield  {title} {\enquote {\bibinfo {title} {{Tailoring
  ${T}_{c}$ by symmetry principles: The concept of superconducting fitness}},}\
  }\href {\doibase 10.1103/PhysRevB.98.024501} {\bibfield  {journal} {\bibinfo
  {journal} {Phys. Rev. B}\ }\textbf {\bibinfo {volume} {98}},\ \bibinfo
  {pages} {024501} (\bibinfo {year} {2018})}\BibitemShut {NoStop}%
\bibitem [{\citenamefont {Brydon}\ \emph {et~al.}(2018)\citenamefont {Brydon},
  \citenamefont {Agterberg}, \citenamefont {Menke},\ and\ \citenamefont
  {Timm}}]{Brydon2018b}%
  \BibitemOpen
  \bibfield  {author} {\bibinfo {author} {\bibfnamefont {P.~M.~R.}\
  \bibnamefont {Brydon}}, \bibinfo {author} {\bibfnamefont {D.~F.}\
  \bibnamefont {Agterberg}}, \bibinfo {author} {\bibfnamefont {Henri}\
  \bibnamefont {Menke}}, \ and\ \bibinfo {author} {\bibfnamefont
  {C.}~\bibnamefont {Timm}},\ }\bibfield  {title} {\enquote {\bibinfo {title}
  {{{B}ogoliubov {F}ermi surfaces: General theory, magnetic order, and
  topology}},}\ }\href {\doibase 10.1103/PhysRevB.98.224509} {\bibfield
  {journal} {\bibinfo  {journal} {Phys. Rev. B}\ }\textbf {\bibinfo {volume}
  {98}},\ \bibinfo {pages} {224509} (\bibinfo {year} {2018})}\BibitemShut
  {NoStop}%
\bibitem [{\citenamefont {Andersen}\ \emph {et~al.}(2019)\citenamefont
  {Andersen}, \citenamefont {Ramires}, \citenamefont {Wang}, \citenamefont
  {Lorenz},\ and\ \citenamefont {Ando}}]{Ramires2019}%
  \BibitemOpen
  \bibfield  {author} {\bibinfo {author} {\bibfnamefont {L.}~\bibnamefont
  {Andersen}}, \bibinfo {author} {\bibfnamefont {A.}~\bibnamefont {Ramires}},
  \bibinfo {author} {\bibfnamefont {Z.}~\bibnamefont {Wang}}, \bibinfo {author}
  {\bibfnamefont {T.}~\bibnamefont {Lorenz}}, \ and\ \bibinfo {author}
  {\bibfnamefont {Y.}~\bibnamefont {Ando}},\ }\href
  {https://arxiv.org/abs/1908.08766} {\enquote {\bibinfo {title} {{Generalized
  Anderson's theorem for superconductors derived from topological
  insulators}},}\ } (\bibinfo {year} {2019}),\ \Eprint
  {http://arxiv.org/abs/1908.08766} {arXiv:1908.08766} \BibitemShut {NoStop}%
\bibitem [{\citenamefont {Scheurer}\ \emph {et~al.}(2015)\citenamefont
  {Scheurer}, \citenamefont {Hoyer},\ and\ \citenamefont
  {Schmalian}}]{Scheurer2015}%
  \BibitemOpen
  \bibfield  {author} {\bibinfo {author} {\bibfnamefont {M.~S.}\ \bibnamefont
  {Scheurer}}, \bibinfo {author} {\bibfnamefont {M.}~\bibnamefont {Hoyer}}, \
  and\ \bibinfo {author} {\bibfnamefont {J.}~\bibnamefont {Schmalian}},\
  }\bibfield  {title} {\enquote {\bibinfo {title} {Pair breaking in
  multiorbital superconductors: An application to oxide interfaces},}\ }\href
  {\doibase 10.1103/PhysRevB.92.014518} {\bibfield  {journal} {\bibinfo
  {journal} {Phys. Rev. B}\ }\textbf {\bibinfo {volume} {92}},\ \bibinfo
  {pages} {014518} (\bibinfo {year} {2015})}\BibitemShut {NoStop}%
\bibitem [{\citenamefont {Mineev}\ and\ \citenamefont
  {Samokhin}(2007)}]{Mineev2007}%
  \BibitemOpen
  \bibfield  {author} {\bibinfo {author} {\bibfnamefont {V.~P.}\ \bibnamefont
  {Mineev}}\ and\ \bibinfo {author} {\bibfnamefont {K.~V.}\ \bibnamefont
  {Samokhin}},\ }\bibfield  {title} {\enquote {\bibinfo {title} {Effects of
  impurities on superconductivity in noncentrosymmetric compounds},}\ }\href
  {\doibase 10.1103/PhysRevB.75.184529} {\bibfield  {journal} {\bibinfo
  {journal} {Phys. Rev. B}\ }\textbf {\bibinfo {volume} {75}},\ \bibinfo
  {pages} {184529} (\bibinfo {year} {2007})}\BibitemShut {NoStop}%
\bibitem [{\citenamefont {Liu}\ \emph {et~al.}(2010)\citenamefont {Liu},
  \citenamefont {Qi}, \citenamefont {Zhang}, \citenamefont {Dai}, \citenamefont
  {Fang},\ and\ \citenamefont {Zhang}}]{Liu2010_topins}%
  \BibitemOpen
  \bibfield  {author} {\bibinfo {author} {\bibfnamefont {C.-X.}\ \bibnamefont
  {Liu}}, \bibinfo {author} {\bibfnamefont {X.-L.}\ \bibnamefont {Qi}},
  \bibinfo {author} {\bibfnamefont {H.~J.}\ \bibnamefont {Zhang}}, \bibinfo
  {author} {\bibfnamefont {X.}~\bibnamefont {Dai}}, \bibinfo {author}
  {\bibfnamefont {Z.}~\bibnamefont {Fang}}, \ and\ \bibinfo {author}
  {\bibfnamefont {S.-C.}\ \bibnamefont {Zhang}},\ }\bibfield  {title} {\enquote
  {\bibinfo {title} {{Model Hamiltonian for topological insulators}},}\ }\href
  {\doibase 10.1103/PhysRevB.82.045122} {\bibfield  {journal} {\bibinfo
  {journal} {Phys. Rev. B}\ }\textbf {\bibinfo {volume} {82}},\ \bibinfo
  {pages} {045122} (\bibinfo {year} {2010})}\BibitemShut {NoStop}%
\bibitem [{\citenamefont {Hashimoto}\ \emph {et~al.}(2013)\citenamefont
  {Hashimoto}, \citenamefont {Yada}, \citenamefont {Yamakage}, \citenamefont
  {Sato},\ and\ \citenamefont {Tanaka}}]{Hashimoto_CuBiSe_2013}%
  \BibitemOpen
  \bibfield  {author} {\bibinfo {author} {\bibfnamefont {T.}~\bibnamefont
  {Hashimoto}}, \bibinfo {author} {\bibfnamefont {K.}~\bibnamefont {Yada}},
  \bibinfo {author} {\bibfnamefont {A.}~\bibnamefont {Yamakage}}, \bibinfo
  {author} {\bibfnamefont {M.}~\bibnamefont {Sato}}, \ and\ \bibinfo {author}
  {\bibfnamefont {Y.}~\bibnamefont {Tanaka}},\ }\bibfield  {title} {\enquote
  {\bibinfo {title} {{Bulk Electronic State of Superconducting Topological
  Insulator}},}\ }\href {\doibase 10.7566/JPSJ.82.044704} {\bibfield  {journal}
  {\bibinfo  {journal} {J. Phys. Soc. Jpn.}\ }\textbf {\bibinfo {volume}
  {82}},\ \bibinfo {pages} {044704} (\bibinfo {year} {2013})}\BibitemShut
  {NoStop}%
\bibitem [{\citenamefont {Balatsky}\ \emph {et~al.}(2006)\citenamefont
  {Balatsky}, \citenamefont {Vekhter},\ and\ \citenamefont
  {Zhu}}]{RMP_impurity}%
  \BibitemOpen
  \bibfield  {author} {\bibinfo {author} {\bibfnamefont {A.~V.}\ \bibnamefont
  {Balatsky}}, \bibinfo {author} {\bibfnamefont {I.}~\bibnamefont {Vekhter}}, \
  and\ \bibinfo {author} {\bibfnamefont {Jian-Xin}\ \bibnamefont {Zhu}},\
  }\bibfield  {title} {\enquote {\bibinfo {title} {Impurity-induced states in
  conventional and unconventional superconductors},}\ }\href {\doibase
  10.1103/RevModPhys.78.373} {\bibfield  {journal} {\bibinfo  {journal} {Rev.
  Mod. Phys.}\ }\textbf {\bibinfo {volume} {78}},\ \bibinfo {pages} {373}
  (\bibinfo {year} {2006})}\BibitemShut {NoStop}%
\bibitem [{\citenamefont {Gastiasoro}\ and\ \citenamefont
  {Andersen}(2018)}]{Gastiasoro2018}%
  \BibitemOpen
  \bibfield  {author} {\bibinfo {author} {\bibfnamefont {Maria~N.}\
  \bibnamefont {Gastiasoro}}\ and\ \bibinfo {author} {\bibfnamefont {Brian~M.}\
  \bibnamefont {Andersen}},\ }\bibfield  {title} {\enquote {\bibinfo {title}
  {Enhancing superconductivity by disorder},}\ }\href {\doibase
  10.1103/PhysRevB.98.184510} {\bibfield  {journal} {\bibinfo  {journal} {Phys.
  Rev. B}\ }\textbf {\bibinfo {volume} {98}},\ \bibinfo {pages} {184510}
  (\bibinfo {year} {2018})}\BibitemShut {NoStop}%
\end{thebibliography}%

\end{document}